\documentclass[a4paper,11pt]{article}
\usepackage{pos}
\usepackage{amsmath}
\usepackage{graphicx}
\usepackage{wrapfig}

\title{New measurement of transverse spin effects in (di-)hadron production from muon-deuteron semi-inclusive DIS at COMPASS}

\author*[a]{Siranush Asatryan}

\onbehalf{for the COMPASS collaboration}

\affiliation[a]{A. Alikhanyan National Science Laboratory,\\
  2 Alikhanyan brothers street, Yerevan, Armenia}

\emailAdd{siranush.asatryan@cern.ch}

\abstract{
The COMPASS experiment is a fixed target high-energy physics experiment that has been collecting data for 20 years (2002 to 2022) at the M2 beamline (SPS, North Area) at CERN. One of the goals of the experiment’s broad physics program was to perform semi-inclusive measurements of target-spin dependent asymmetries in (di-)hadron production in deep inelastic scattering (DIS) with high-energy muons scattering off polarized nucleons. The latest COMPASS semi-inclusive DIS measurements were performed in 2022 using a transversely polarized $^6$LiD target and a 160 GeV/c muon beam. The first results from approximately two-thirds of the new data exhibit total uncertainties that are up to a factor of three smaller than those observed in previous deuteron measurements. These results serve to balance the existing data collected with transversely polarized proton targets, which is of particular importance in constraining the d-quark transversity and Sivers functions.
}

\FullConference{%
  31st International Workshop on Deep Inelastic Scattering\\
 8 - 12 April, 2024\\
  Maison MINATEC, Grenoble, FRANCE
}

\begin{document}
\maketitle

\section{Introduction}




The semi-inclusive measurements of hadron production in deep inelastic lepton-nucleon scattering, $\ell \,N \rightarrow \ell^\prime \,h \, X$ (hereafter referred to as SIDIS), play a significant role in the study of the three-dimensional parton structure of nucleons in momentum space. 

\begin{wrapfigure}{r}{0.35\textwidth}
    \centering
    \includegraphics[width=0.9\linewidth]{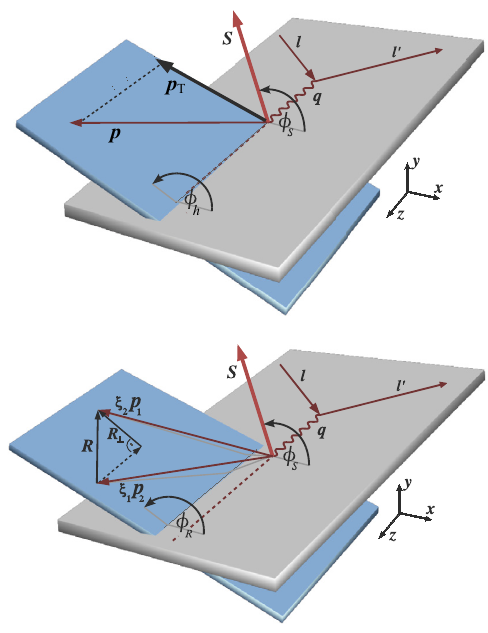}
    \caption{SIDIS framework and notations: top - single hadron production, bottom - dihadron production.}
    \label{fig:frame}
\end{wrapfigure}

The SIDIS cross section can be factorized into convolutions of perturbatively calculable hard-scattering parton cross sections and two categories of nonperturbative functions. The latter are the parton distribution functions (PDFs), which describe the distribution of quarks in the target nucleon, and the fragmentation functions (FFs), which describe the hadronization of a quark into the observed hadron~\cite{Collins:2016hqq}.

The three-dimensional picture of nucleon encompasses both the longitudinal and intrinsic transverse motion of partons within (un)polarized hadrons, as well as their spin degrees of freedom. Within the leading-twist (twist-2) approximation of pQCD, there exist eight parton transverse-momentum-dependent (TMD) PDFs of the nucleon, which describe the distributions of longitudinal and transverse momenta of partons and their correlations with nucleon and quark spins. 
The TMD PDFs are accessed via measurements of specific azimuthal asymmetries in SIDIS (for recent reviews see \textit{e.g.} Refs.~\cite{Anselmino:2020vlp,Avakian:2019drf}).
Two of the most extensively studied TMD PDFs are the transversity function, denoted as $h_1$, and the Sivers function, denoted as $f_{1T}^\perp$.

For each given quark flavor $h_1$ function encodes the description of transversely polarized quarks within a transversely polarized nucleon. The transversity PDF is chiral odd, which implies that it can only be accessed in conjunction with another chiral-odd function. The corresponding azimuthal asymmetry appearing in SIDIS is referred to as the Collins asymmetry $A^h_{\rm{Coll}}$.
In the leading order of perturbative QCD, this asymmetry is proportional to the convolution over transverse momenta of the transversity function and the so-called Collins fragmentation function~\cite{Kotzinian:1994dv,Bacchetta:2006tn,Bastami:2018xqd}.
The Collins FF $H_{1q}^{\perp h}$, describes the correlation between the transverse spin of the quark $q$ and the transverse momentum of the unpolarized final-state hadron $h$.

The Collins asymmetry $A^h_{\rm{Coll}}$ is characterized by a modulation
$[1 + D_{\rm{NN}} f P_t \sin \phi_{\rm{Coll}}]$, in the number of final state hadrons, where $\phi_{\rm{Coll}} = \phi_h + \phi_S - \pi$ is the Collins angle, and $\phi_h$ and
$\phi_S$ are the azimuthal angles of the hadron and of the target nucleon spin vector, respectively~\cite{COMPASS:2023vhr} (see Figure~\ref{fig:frame}).
In the amplitude of the modulation, $D_{\rm{NN}}\simeq(1 - y)/(1 - y + y^2/2)$ is the depolarization coefficient, $y$ represents the fractional energy of the virtual photon, $f$ is the dilution factor of the target material and $P_t$ denotes the transverse polarization of the target nucleon.

An alternative method to access the transversity PDF is to perform SIDIS measurements of dihadron (oppositely charged hadron pair, $h^{+}h^{-}$) production with both hadrons produced within the current fragmentation region~\cite{Radici:2001na,COMPASS:2014ysd}. 
In this case the transversity PDF is paired with a new chiral-odd FF, the dihadron Fragmentation Function (DiFF) $H_{1}^{\sphericalangle}$, which describes the spin-dependent part of the fragmentation of a transversely polarized quark into a pair of unpolarized hadrons. This gives rise to the $A_{UT}^{\sin(\phi_R-\phi_S-\pi)}$ asymmetry, where $\phi_R$ is the dihadron azimuthal angle~\cite{COMPASS:2014ysd} (see. Figure~\ref{fig:frame}). The analysis of dihadron SIDIS measurements can be performed using collinear factorization, as demonstrated in Ref.~\cite{Bacchetta:2008wb}.

The Sivers TMD PDF $f_{1T}^\perp$ describes the correlation of the intrinsic transverse momentum $k_T$ of unpolarized quarks with the transverse spin of the nucleon. The Sivers asymmetry $A^h_{Siv}$ is proportional to the convolution of $f_{1T}^\perp$ and the spin-averaged fragmentation function $D_{1q}^{h}$ and is given by the modulation $[1 + f P_t \sin \phi_{\rm{Siv}}]$ in the distribution of the hadrons produced in SIDIS off transversely polarized nucleons~\cite{Bacchetta:2006tn,Bastami:2018xqd}. Here, $\phi_{\rm{Siv}} = \phi_h - \phi_S$ is the Sivers angle.

The first COMPASS measurements of the Sivers and Collins effects with a deuteron target were made from 2002 to 2004. The results indicated that there are no significant effects, with the measurements being compatible with zero across the entire kinematic range, albeit with relatively large uncertainties~\cite{COMPASS:2006mkl,COMPASS:2008isr}. 
It was the only existing transverse deuteron target data until 2022.
Subsequently, in 2007 and 2010, COMPASS performed SIDIS measurements with a transversely polarized proton (NH$_3$) target, achieving significantly enhanced precision. The Collins and Sivers asymmetries exhibit a clear non-zero behaviour for a proton target and were found to be consistent with the results previously obtained by the HERMES experiment~\cite{COMPASS:2014bze,HERMES:2009lmz,HERMES:2010mmo}. These results implied the existence and significance of the TMD effects in the context of nucleon spin structure studies.
COMPASS data are used continuously in global phenomenological fits conducted by several research groups with the objective of extracting transversity and Sivers PDFs~\cite{Anselmino:2015sxa,Radici:2018iag,Gamberg:2022kdb}. The separation of quark flavors can be achieved through the use of proton and deuteron (or neutron) targets, as evidenced by the results of the aforementioned measurements. The lack of precision in the available deuteron data, has resulted in a poorly constrained d-quark transversity PDF. This limitation prompted the new COMPASS measurements with a transversely polarized deuteron target. The aim was to significantly increase the available statistics in order to balance the existing proton data and reduce the uncertainties of both $u$ and $d$ quark TMD PDFs~\cite{compass-ii}.

\section{New COMPASS deuteron measurements and data analysis}
For a comprehensive overview of the COMPASS apparatus, measurement principles, and data analysis, please refer to the Refs.~\cite{COMPASS:2023vhr,COMPASS:2012ozz,COMPASS:2012dmt}. A brief summary is provided here for the sake of completeness.  
The COMPASS spectrometer was situated at the M2 beamline of the CERN SPS ~\cite{COMPASS:2007rjf}. In 2022, the spectrometer configuration was similar to that employed in 2007 and 2010~\cite{COMPASS:2012ozz,COMPASS:2012dmt}, when SIDIS measurements with a transversely polarized proton target were conducted.
In particular, compared to the 2002-2004 measurements the angular acceptance was enlarged from 70~mrad in 2002-2004 to 180~mrad.
The data were collected by scattering a 160 GeV/$c$ muon beam on a $^6$LiD target, consisting of three cylindrical cells with a total length of 120 cm. The neighbouring cells were polarized in opposite vertical directions, thereby allowing the collection of data for both spin directions simultaneously. The data collection was performed in approximately two week long periods. At the midpoint of each period, the spin orientation within the target was reversed to further minimize systematic effects.

In the data analysis, standard DIS selections are applied: only events with photon virtuality $Q^2 > 1$ GeV/$c^2$, $0.1 < y < 0.9$ and mass of the hadronic final state system $W > 5$ GeV/$c^2$ are considered. 
Hadron observables are additionally required to satisfy the following criteria: the transverse momentum with respect to the virtual photon direction of $p_T > 0.1$ GeV/$c$ and a fraction of the available energy of $z > 0.2$.
This resulted in a total of approximately 40 million positive hadrons and 32 million negative hadrons.
For dihadrons additional criteria are applied: $z$ and Feynman $x_F$ are required to be above 0.1 for both hadrons, the misisng energy of the system is required to be above 3 GeV/$c$ to suppress exclusive events and relative transverse momentum of the pair is required to be $R_T > 0.07$ GeV/$c$.
An extended unbinned maximum likelihood estimator~\cite{COMPASS:2010hbb} is employed to simultaneously extract all the azimuthal asymmetries in the transverse spin-dependent part of the SIDIS (di)hadron production cross section~\cite{Bacchetta:2006tn,Bastami:2018xqd}.
In order to extract the Collins and Sivers asymmetries, the measured amplitudes of the modulations in $\sin{\phi_{Coll}}$ and $\sin{\phi_{Siv}}$ are corrected by the factors $f$, $P_t$, and $D_{NN}$ (only Collins).

\section{Results}

The following presentation of results is based on data from seven out of ten periods, which corresponds to approximately two-thirds of the total statistics collected in 2022~\cite{COMPASS:2023vhr}.

\begin{figure}[h]
    \centering
    \includegraphics[width=0.49\linewidth]{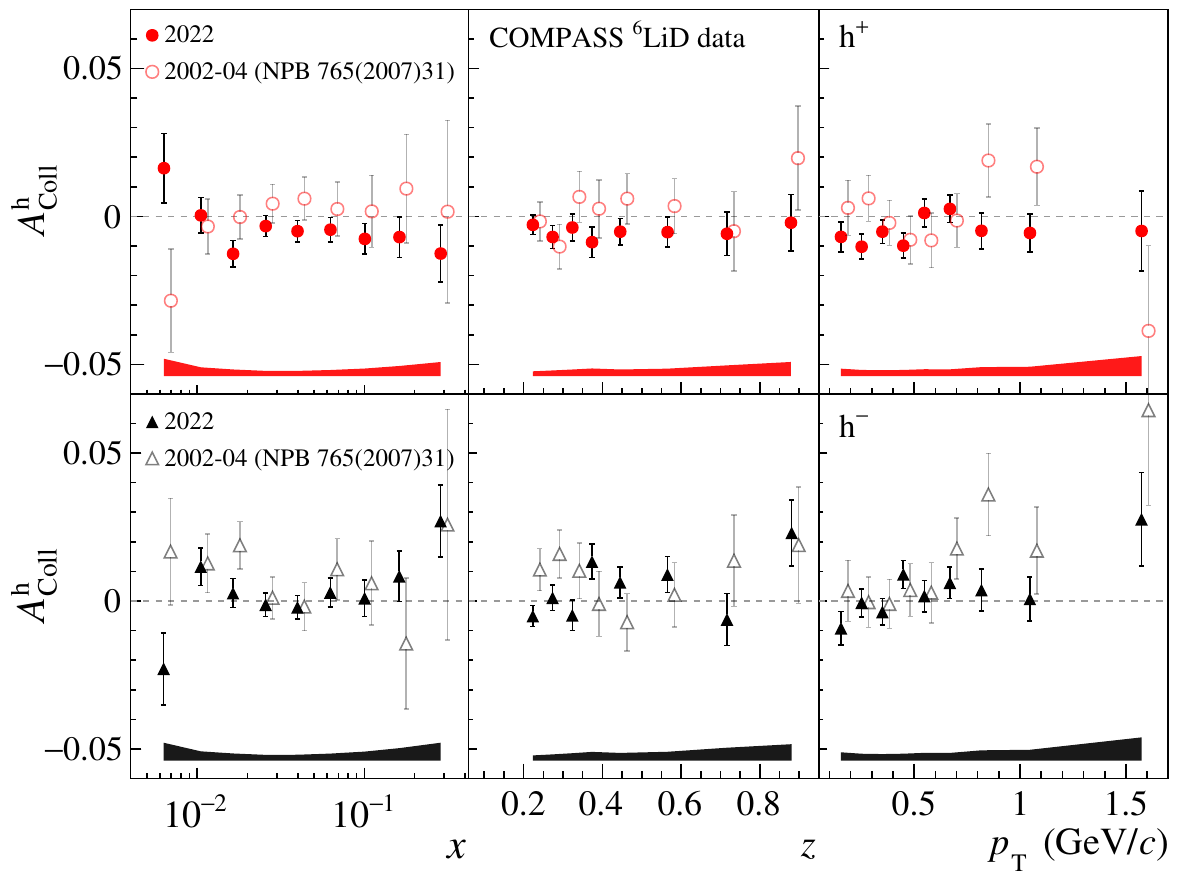}
    \includegraphics[width=0.49\linewidth]{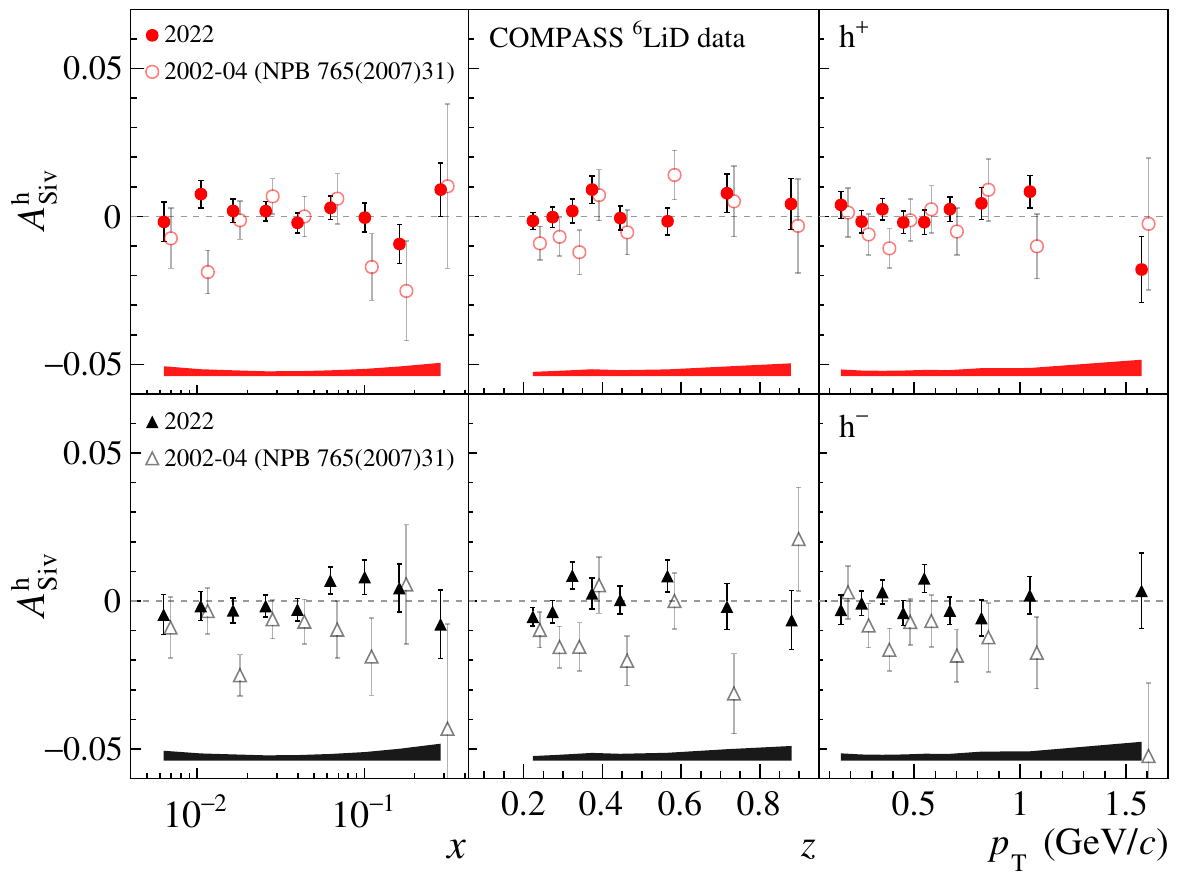}
    \caption{The Collins (left) and Sivers (right) deuteron asymmetries for positive (red circles) and negative (black triangles) hadrons. The results are from the 2022~\cite{COMPASS:2023vhr} (filled markers) and the 2002-2004 data sets~\cite{COMPASS:2006mkl} (empty markers). The 2022 error bars are statistical only, the bands show the systematic uncertainties.}
    \label{fig:1hadron}
\end{figure}

\begin{figure}[h]
    \centering
    \includegraphics[width=0.7\linewidth]{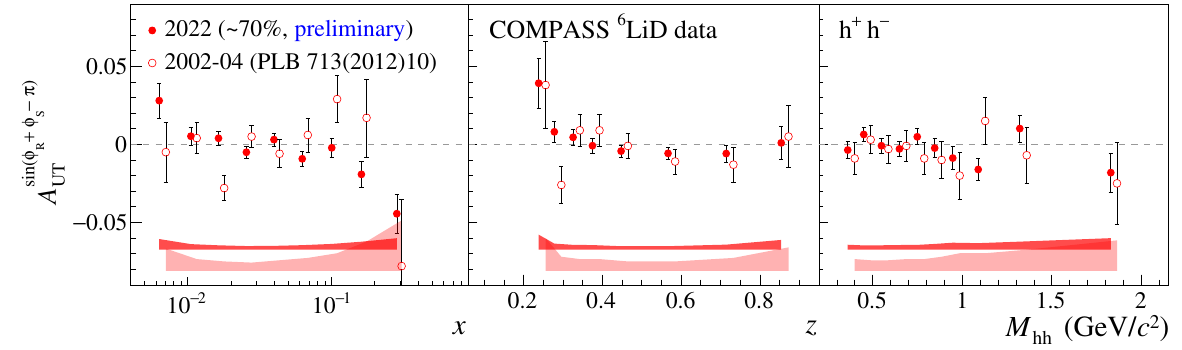}
    \includegraphics[width=0.7\linewidth]{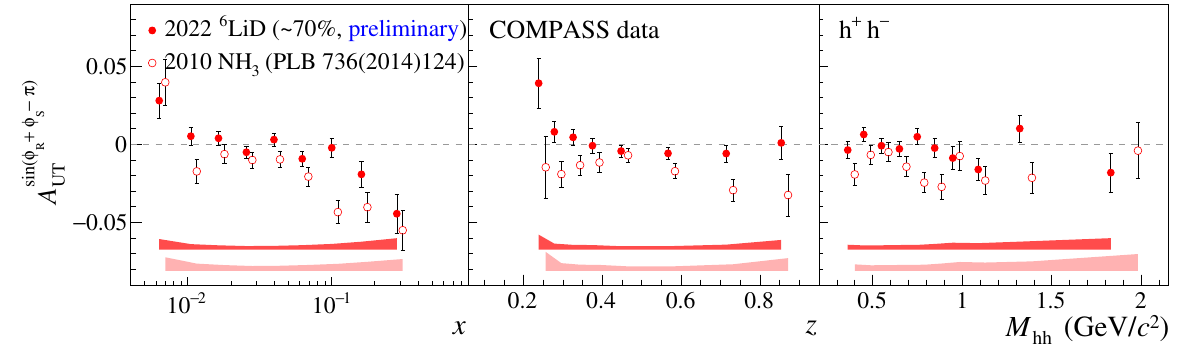}
    \caption{New preliminary deuteron 2022 results for Collins asymmetries are compared with previous 2002-2004 COMPASS deuteron data~\cite{COMPASS:2014ysd} (top plot) and with the COMPASS proton data~\cite{COMPASS:2012bfl} (bottom plot).}
    \label{fig:dihadron}
\end{figure}

\begin{figure}[h]
    \centering
    \includegraphics[width=0.39\linewidth]{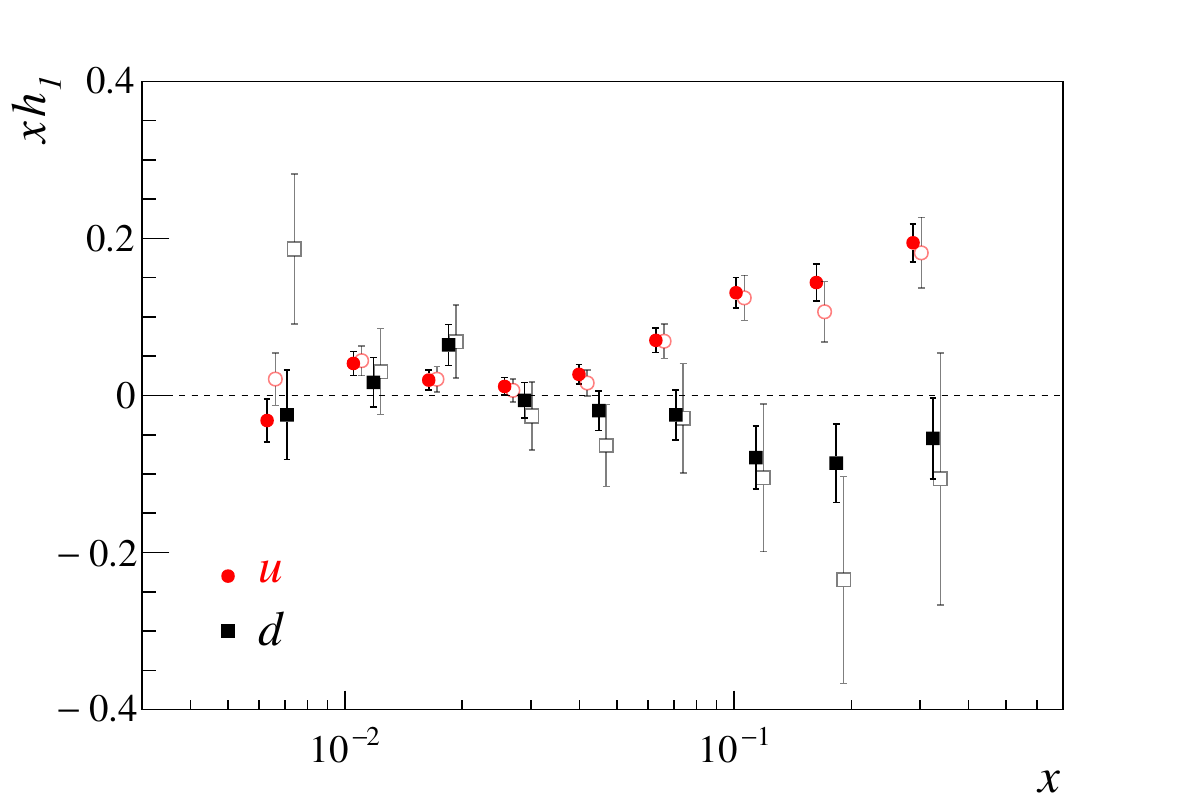}
    \includegraphics[width=0.39\linewidth]{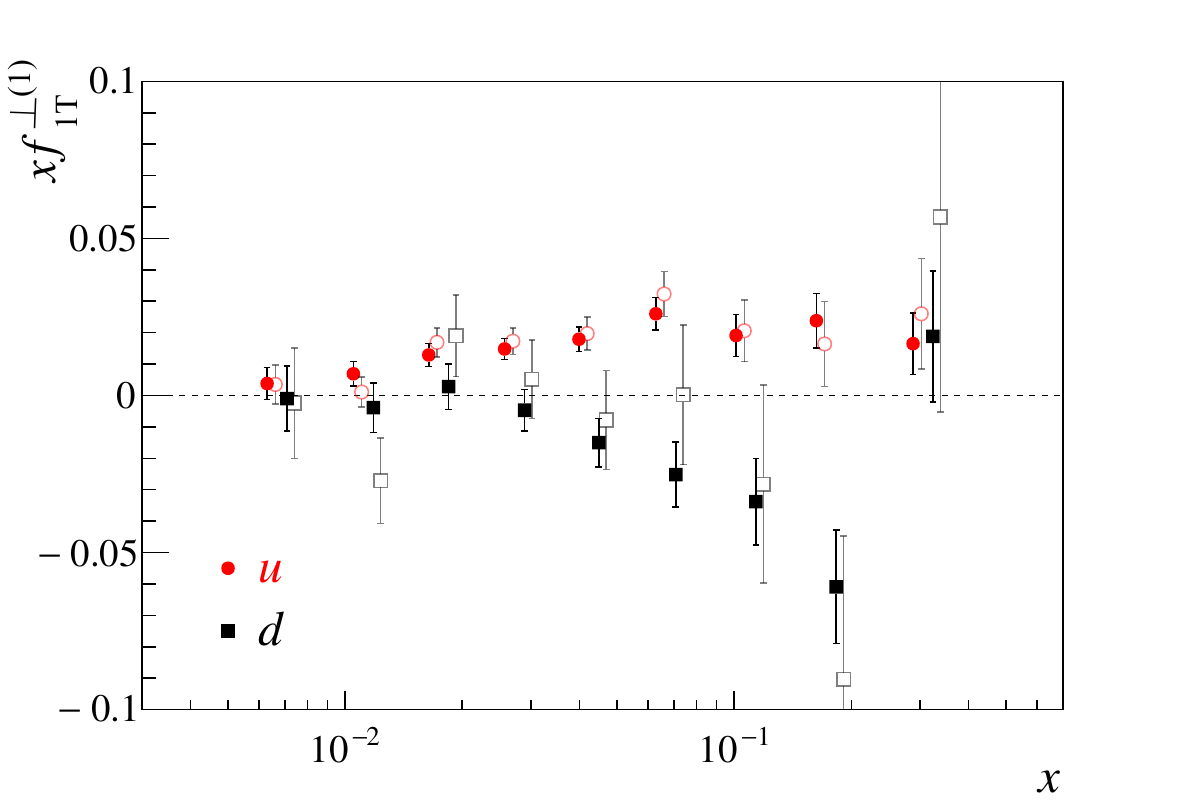}
    \caption{Left panel: the valence transversity functions for the $u$ (red circles) and $d$ (black squares) quarks. The values were obtained using the previously published results for the proton and deuteron Collins asymmetries (open points) and including the 2022 deuteron results~\cite{COMPASS:2023vhr} (filled points). The error bars illustrate the statistical uncertainties. Right panel: the same data for the first $k_T^2$ moments of the Sivers functions.}
    \label{fig:functions}
\end{figure}

Figure~\ref{fig:1hadron} shows the results for the $A^h_{\rm{Coll}}$ (right panel) and $A^h_{\rm{Siv}}$ (left panel) asymmetries in bins of $x$, $z$ and $p_T$ for both positive and negative hadrons. The new asymmetries from the 2022 measurement~\cite{COMPASS:2023vhr} are compared with the COMPASS data from 2002-2004~\cite{COMPASS:2005csq,COMPASS:2006mkl,COMPASS:2008isr}.
The two sets of results are found to be in agreement, with a notable gain in precision in 2022 (the uncertainties are reduced by up to a factor of three). 

The present measurement of the Collins asymmetry indicates the presence of small negative (positive) signals for positive (negative) hadrons at large $x$. Same is true for the dihadron Collins asymmetries shown in Figure~\ref{fig:dihadron}. All Collins asymmetries exhibit the same sign as those previously observed with the proton target. As for the Sivers asymmetries, they remain compatible with zero despite the reduction in uncertainties.

Figure~\ref{fig:functions} illustrates the results of a point-by-point extraction of the transversity and Sivers PDFs, employing the framework introduced in Refs.~\cite{Martin:2014wua,Martin:2017yms}. The enhanced precision is apparent, particularly for the $d$ quark transversity PDF, with a more precise determination of its sign.

\section{Conclusion}

The first results from high-statistics COMPASS measurements performed in 2022 with a transversely polarized deuteron target were reported. The current analysis is based on approximately two-thirds of the total statistics collected in 2022.
The overall precision of the Collins and Sivers measurements has been increased by up to a factor of three in comparison to the previously published COMPASS deuteron results~\cite{COMPASS:2006mkl,COMPASS:2014ysd}. This improvement is particularly evident in the point-to-point extractions of the transversity and Sivers functions for $u$ and $d$ valence quarks, which clearly demonstrate the effect at the level of PDFs.  

In conclusion, the new COMPASS deuteron results are expected to have a profound impact on the global fits of the transversity and Sivers TMD PDFs, thereby enhancing our understanding of the transverse-spin structure of the nucleon.

\acknowledgments

The work of S.A. was supported by the Higher Education and Science Committee of RA, in the frame of the research project No 21AG-1C028.

\bibliography{ColSiv_biblio}{}

\end{document}